\documentclass[twocolumn]{aa}
\usepackage{graphicx}
\usepackage{amssymb}
\begin{document}
   \title{The sub-damped Ly\,$\alpha$ system toward HE\,0001$-$2340: \\
          galaxy formation at $z\approx2$
	     \thanks{Based on observations carried out at the European
	     Southern Observatory (ESO), La Silla, under prog.\,ID No.\,
	     166.A-0106 with the UVES spectrograph at the ESO Very Large
	     Telescope, Paranal, Chile.}}



   \author{P. Richter,
          \inst{1}
          C. Ledoux,
          \inst{2}
          P. Petitjean,
          \inst{3}
	  \and
          J. Bergeron,
	  \inst{3}
          }

   \offprints{P. Richter\\
   \email{prichter@astro.uni-bonn.de}}

   \institute{Institut f\"ur Astrophysik und Extraterrestrische Forschung,
              Auf dem H\"ugel 71, 53121 Bonn, Germany
         \and
              European Southern Observatory, Alonso de C\a'ordova 3107,
	      Casilla 19001, Vitacura, Santiago, Chile
         \and
              Institut d'Astrophysique de Paris - CNRS, 98bis Boulevard Arago, 
              75014, Paris, France
            }

   \date{Received xxx; accepted xxx}

\abstract{
We present a detailed analysis of chemical abundances in a sub-damped Ly\,$\alpha$
absorber (sub-DLA) at $z=2.187$ towards the quasar HE\,0001$-$2340 ($z_{\rm em}=2.28$).
Our study is based on high-resolution ($R\approx 45,000$) spectroscopic data from the UVES
instrument installed on the ESO {\it Very Large Telescope} (VLT).
This sub-DLA system consists of at least 25 individual subcomponents spanning 
a restframe velocity range of $\sim 400$ km\,s$^{-1}$. 
The total neutral hydrogen column density is log $N($H\,{\sc i}$)\approx19.7$.
Detected species include C\,{\sc ii}, 
C\,{\sc iv}, N\,{\sc i}, N\,{\sc ii}, O\,{\sc i},
Mg\,{\sc ii}, Al\,{\sc ii}, Al\,{\sc iii}, 
Si\,{\sc ii}, Si\,{\sc iv}, P\,{\sc ii}, Fe\,{\sc ii}, and possibly D\,{\sc i}.
For the dominating neutral gas component at $v_{\rm rel}=+13$ km\,s$^{-1}$ 
(relative to $z=2.187$) we 
derive an oxygen abundance of [O/H]$=-1.81\pm0.07$ (1/65 solar).
With its extremely low nitrogen content ([N/H]\,$\lesssim -3.3$ and [N/O]\,$\lesssim -1.5$)
the absorber exhibits a classic massive star abundance pattern. 
Our measurements place the $z=2.187$ absorber towards HE\,0001$-$2340 among the systems
with the lowest ever measured [N/$\alpha$] ratios in the Universe. 
The low [N/O] value  
is consistent with the idea that primary nitrogen production by the very first stars 
have enriched the intergalactic gas to a level of [N/O]$\approx-1.5$.
Peculiar abundances are found in the outermost blue components
near $-350$ km\,s$^{-1}$ (in the $z=2.187$ restframe) 
where we observe significant overabundances
of phosphorus ([P/C]$\approx+1.5$), silicon ([Si/C]$\approx+0.6$), and
aluminum ([Al/C]$\approx+0.5$) after correcting for the effects of ionization.
Our study suggests that
the sight line passes through the gaseous environment
of one or more stellar clusters that locally enriched their
interstellar neighbourhood by supernova ejecta
generating the observed abundance anomalies. 
The large velocity spread of the entire
absorption system points to a merger that triggers
the formation of these clusters.
We thus may be observing a young galaxy at $z\approx2$ that 
currently forms out of a merger event.

 
}
\titlerunning{The sub-DLA towards HE\,0001$-$2340}

\maketitle
%

\begin{figure*}[t!]
\resizebox{1.0\hsize}{!}{\includegraphics{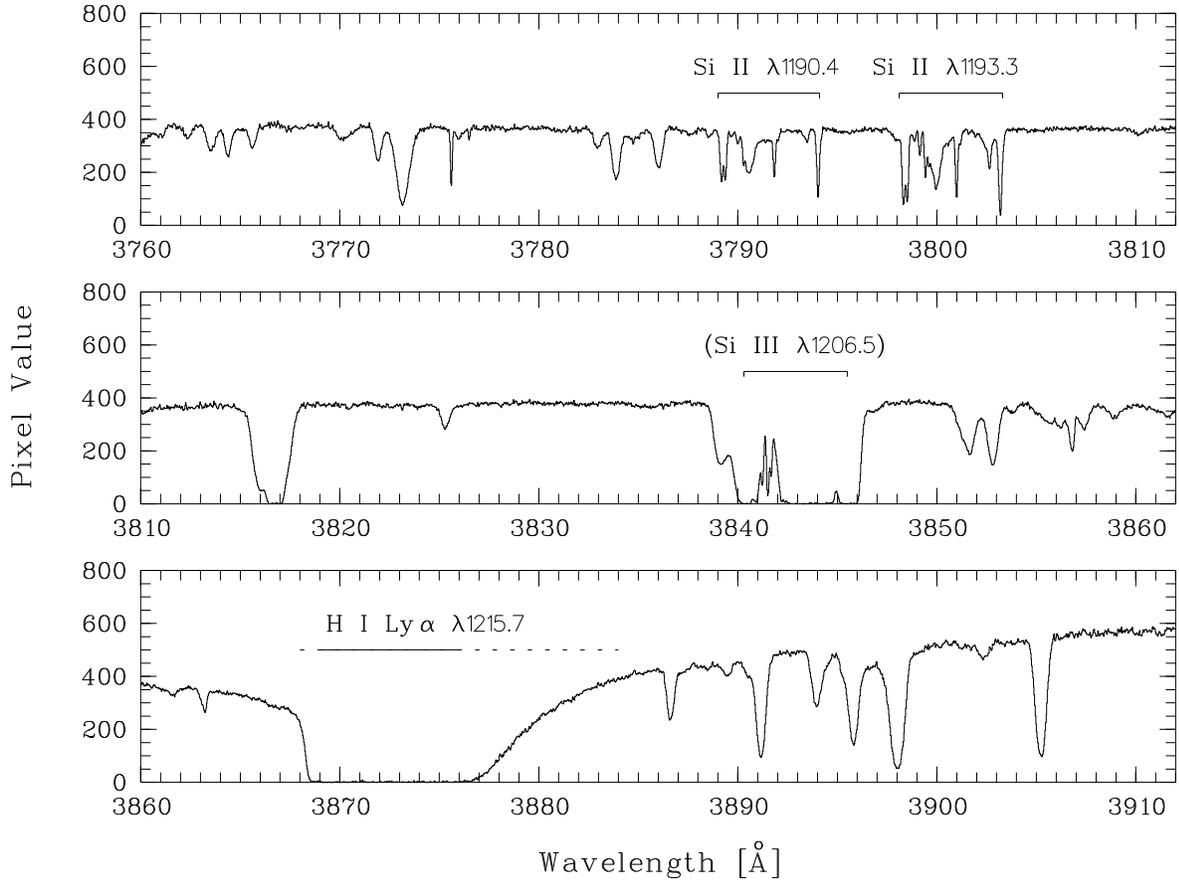}}
\caption[]{
Portion of the UVES spectrum of HE\,0001$-$2340 in the
wavelength range between $3760$ and $3912$ \AA. Absorption
by Si\,{\sc ii} $\lambda \lambda 1190.4, 1193.3$,
Si\,{\sc iii}$ \lambda 1206.5$ (blended by intervening H\,{\sc i}),
and H\,{\sc i} Ly\,$\alpha$ $\lambda 1215.7$ from the
$z=2.187$ sub-DLA is shown in the upper, middle, and lower panel,
respectively.
}
\end{figure*}

\section{Introduction}

Absorption line systems at high redshift provide a wealth 
of information about the chemical evolution of the early
Universe. 
Next to the Ly\,$\alpha$ forest and the 
Lyman limit systems that trace the low-density IGM
or the intergalactic environment 
of galaxies at neutral hydrogen 
column densities of
$N($H\,{\sc i}$) < 10^{19}$ cm\,$^{-2}$,
the physical and chemical properties 
of absorbers with $N($H\,{\sc i}$)\geq 10^{19}$ cm\,$^{-2}$
are particularly interesting. These systems 
contain most of the neutral gas mass in the Universe
at $z>1$ (Lanzetta, Wolfe, \& Turnshek 1995; Wolfe et al.\,1995) and 
consist of heavy elements in an abundance pattern that 
suggests the chemical enrichment by the first generations
of stars (e.g., Lu et al.\,1996; Prochaska
\& Wolfe 1999). Such absorbers therefore may
represent progenitors of present-day galaxies.

Most of the information about abundances of 
metals, dust, and molecules in these systems has 
been obtained from the damped Ly\,$\alpha$ systems (DLAs),
which have H\,{\sc i} column densities $> 2\times 10^{20}$ 
cm\,$^{-2}$ (e.g., Ledoux, Petitjean, \& Srianand\,2003). 
Absorption systems with somewhat lower
column densities ($10^{19}$ cm\,$^{-2} \leq N($H\,{\sc i}$) 
\leq 2\times 10^{20}$ cm\,$^{-2}$) are the so-called sub-damped
Ly\,$\alpha$ systems (sub-DLAs; e.g., Dessauges-Zavadsky et al.\,2003;
P\a'eroux et al.\,2003). While there is no
compelling evidence that DLAs and sub-DLAs trace
different classes of protogalactic objects, abundance
measurements in sub-DLAs are of particular interest
because of their absorption characteristics.
For total H\,{\sc i} column densities
$\leq 2\times 10^{20}$ cm\,$^{-2}$ and metallicities
less than $\sim 0.1$ solar, the absorption
lines from neutral oxygen and neutral nitrogen 
(two important elements for disentangling the complex chemical
enrichment history of high-$z$ absorbers)
are relatively weak.
This enables us
to derive accurate abundances for these two important
elements in sub-DLAs without facing the problem of heavy
line saturation in the oxygen lines.

In this paper we present the detailed analysis of a 
sub-damped Ly\,$\alpha$ system at $z=2.187$ in the
direction of the quasar HE\,0001$-$2340. This paper 
represents the first one in a series of two dealing
with the chemical and physical properties of particularly 
interesting sub-DLA systems in the data set of 
the ESO-VLT Large Programme
``The Cosmic Evolution of the IGM''.
This paper is organized as follows:
in Sect.\,2 we present the observations, the data reduction, 
and the data analysis method. 
Absorption by neutral and weakly ionized species is presented
in Sect.\,3. Sect.\,4 deals with the properties of
the highly ionized gas. We discuss the results
in Sect.\,5 and summarize our study in Sect.\,6.

\section{Observations, data handling, and analysis method}

The quasar HE\,0001$-$2340 
($V=16.7$, $z_{\rm em}=2.28$) was observed with
the UVES instrument installed on the VLT
as part of the ESO-VLT Large Programme
``The Cosmic Evolution of the IGM'', which aims
at providing a homogeneous data set of high-resolution
($R\sim45,000$), high signal-to-noise (S/N$>30$)
spectral data for QSO absorption line spectroscopy of the IGM
in the redshift range between $z=2.2-4.5$.

HE\,0001$-$2340 was observed through a 1'' slit
(with a seeing typically $\leq0.8$'') with two
setups using dichroic beams splitters for the 
blue and the red arm (Dic\,1, B346nm$+$R580nm, and
Dic\,2, B437nm$+$860nm, respectively). 
This setup provides a wavelength coverage
from $\sim 3050$ to $10,400$ \AA, with small 
gaps near $5750$ and $8550$ \AA. 
The total integration time (Dic\,1 $+$ Dic\,2) for 
HE\,0001$-$2340 was 12 hours.
The raw data were reduced using the UVES pipeline 
implemented in the ESO-MIDAS software package.
The pipeline reduction includes flat-fielding, 
bias- and sky-subtraction, and a relative
wavelength calibration. The individual spectra 
then have been 
coadded and corrected to vacuum wavelengths.
The S/N in the spectrum generally is very high and
attains a maximum of $\sim 150$ per resolution
element for $\lambda > 4000$ \AA.
In Fig.\,1, a representative portion of the UVES spectrum of 
HE\,0001$-$2340 in the wavelength range between
$3760$ and $3912$ \AA\, is displayed,
showing selected absorption lines of the 
sub-DLA system at $z=2.187$.

\begin{figure}[t!]
\resizebox{1.0\hsize}{!}{\includegraphics{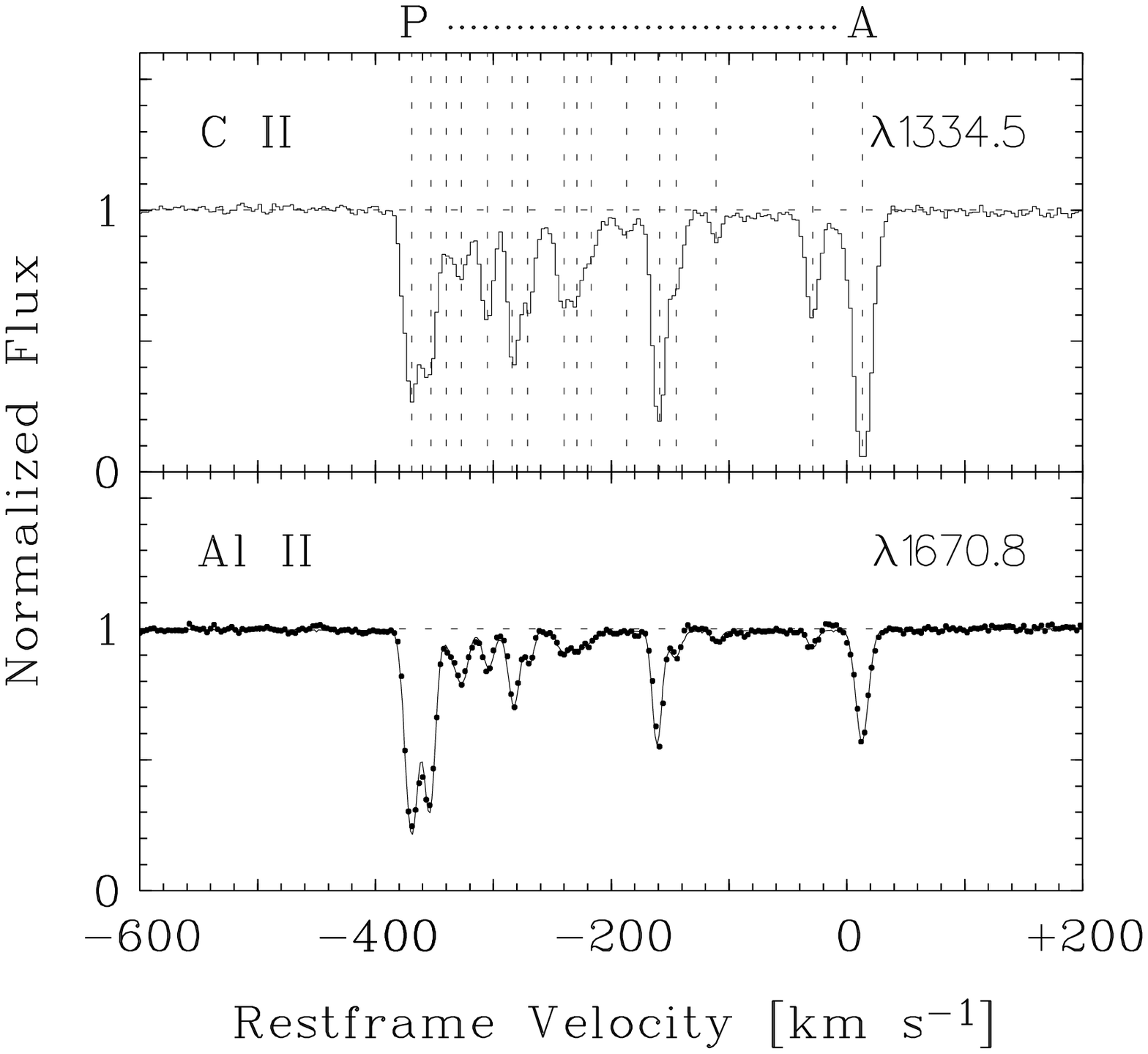}}
\caption[]{
Absorption profiles of C\,{\sc ii} and Al\,{\sc ii} from
the $z=2.187$ sub-DLA are displayed (velocities refer
to the $z=2.187$ restframe). 
The C\,{\sc ii} $\lambda 1334.5$ absorption (upper panel)
demonstrates the complex velocity structure of this absorber.
16 individual components are identified, given by the letters
A to P. These components are indicated by 
the vertical dashed lines. The lower panel
shows an example for a multi-component Voigt profile fit from
the {\tt FITLYMAN} routine.  Absorption by Al\,{\sc ii} 
$\lambda 1670.8$ (dots) is overlayed by the 
corresponding optimum fit (solid line).
}
\end{figure}

For the spectral analysis of the data we made use of the
{\tt FITLYMAN} package in MIDAS (Fontana \& Ballester 1995).
This routine uses a $\chi ^2$ minimization algorithm 
to derive column densities, $N$, and Doppler-parameters, $b$, 
via Voigt-profile fitting, taking into account
the spectral resolution of the instrument (see above).
First of all, we have disentangled 
the component structure in the $z=2.187$ absorber for a)
neutral and weakly ionized species (singly ionized species plus
Al\,{\sc iii}), and b) highly ionized
species (C\,{\sc iv} and Si\,{\sc iv}), as explained
in detail in Sects.\,3 and 4. 
Using a number of high S/N absorption
profiles from various species we then 
obtained Doppler parameters for each component by
Voigt profile fitting. We 
use a single $b$ value per absorption component to fit 
all neutral and weakly ionized species except 
hydrogen and deuterium. 
Fig.\,2, lower panel, shows 
an example for a 
Voigt profile fit to
the Al\,{\sc ii} $\lambda 1670.8$ line. The
spectral data are plotted with black points, the 
corresponding fit is shown as solid line.
Throughout this paper we use wavelengths and 
oscillator strengths 
from the atomic line list of Morton (2003).

\section{Neutral and weakly ionized species}

\subsection{Velocity structure and $b$ values}

\begin{figure*}[t!]
\resizebox{0.68\hsize}{!}{\includegraphics{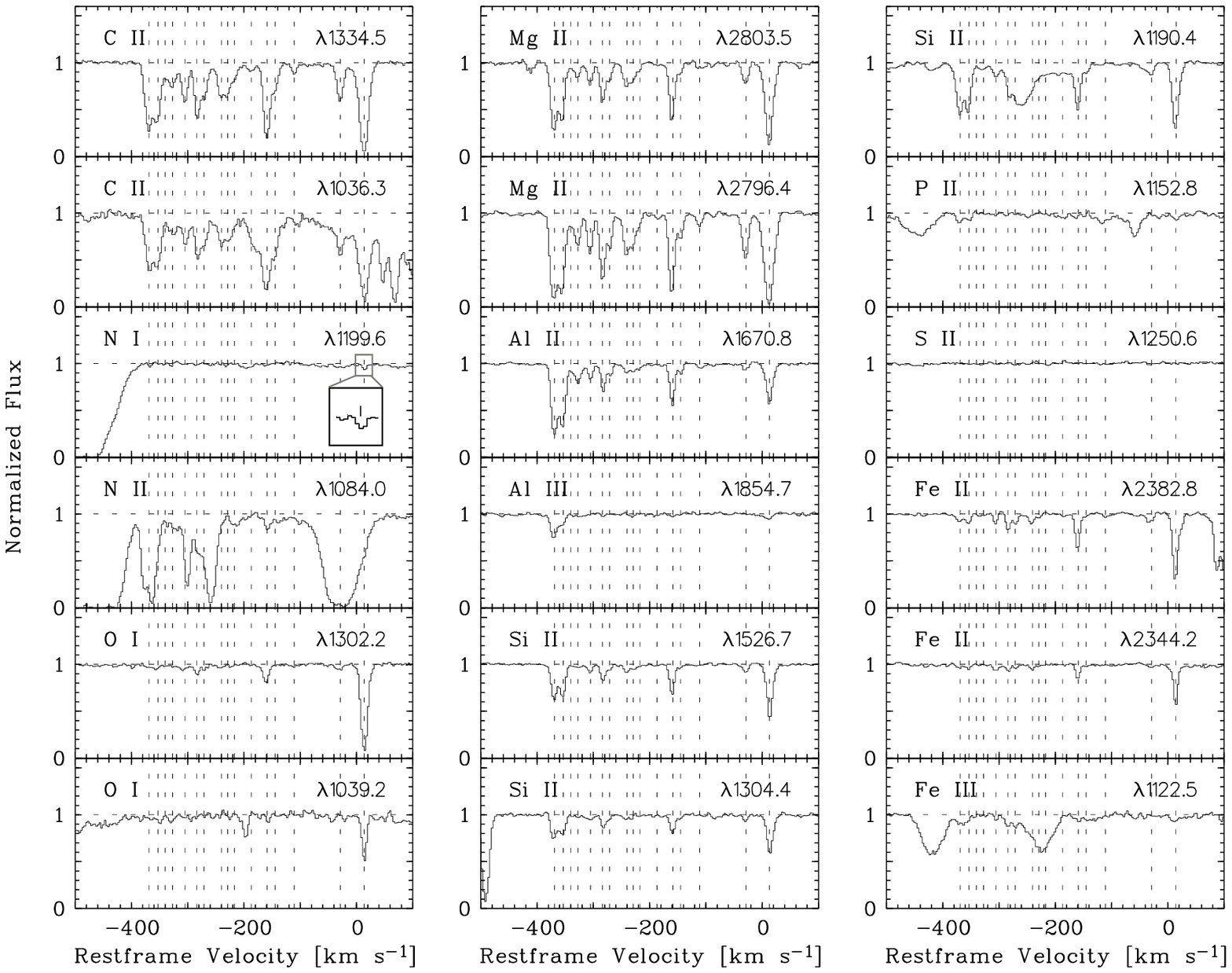}}
\caption[]{
Absorption profiles of C\,{\sc ii}, N\,{\sc i}, N\,{\sc ii}, O\,{\sc i},
Mg\,{\sc ii}, Al\,{\sc ii}, Al\,{\sc iii}, Si\,{\sc ii},
P\,{\sc ii}, S\,{\sc ii}, Fe\,{\sc ii}, and Fe\,{\sc iii} 
for the $z=2.187$ sub-DLA are shown. The
individual absorption components are indicated with
the vertical dashed lines. The velocity scale refers to the $z=2.187$
restframe.
}
\end{figure*}

Fig.\,2 (upper panel) shows the C\,{\sc ii} 
$\lambda 1334.5$ line plotted on 
the $z=2.187$ restframe velocity scale, demonstrating the
complex component structure of this absorption system.
The C\,{\sc ii} line, among others, is sufficiently
strong to show also weak velocity 
components and lies in a region that is
not affected by line blending. Moreover, this line has
high S/N, so that the various subcomponents 
can be reliably identified. To derive a unique model
for the velocity component structure we have fitted Voigt profiles 
to a number of high S/N lines 
of C\,{\sc ii}, Mg\,{\sc ii}, and Si\,{\sc ii}.
We identify $16$ velocity subcomponents,
in Fig.\,2 shown
with vertical dashed lines.
In the $z=2.187$ restframe
these absorption components have relative velocities of
$+13$, $-29$, $-111$,
$-145$, $-159$, $-187$, $-217$, $-229$, 
$-240$, $-271$, $-284$, $-305$,
$-327$, $-340$, $-353$, and $-369$ km\,s$^{-1}$. 
This velocity model then has been used to derive $b$ values and column
densities for each component via Voigt-profile fitting.
The resulting values for
$b$ and log $N$ and their $1\sigma$ errors are listed in Table 1.
Throughout the following, we label
the individual velocity components with the letters A to P in alphabetical order
(i.e, component A is the one at $+13$ km\,s$^{-1}$). The 
strongest absorption in C\,{\sc ii} and other 
neutral and weakly ionized species is seen in
component A, thus at the outermost red wing of the overall
absorption structure. 

\subsection{Metals}

A number of neutral and weakly ionized species are detected 
in the various velocity components, including C\,{\sc ii}, N\,{\sc i}, N\,{\sc ii},
O\,{\sc i}, Mg\,{\sc ii}, Al\,{\sc ii}, Al\,{\sc iii}, Si\,{\sc ii},
P\,{\sc ii}, and Fe\,{\sc ii}. A selection of velocity profiles of these 
species are shown in Fig.\,3. Column densities for these ions
as well as upper limits for undetected species (e.g.,
Zn\,{\sc ii}, S\,{\sc ii}, and molecular hydrogen, H$_2$)
are listed in Table 1.
The strong lines of C\,{\sc ii}
($\lambda\lambda 1334.5, 1036.3$) and Mg\,{\sc ii} ($\lambda\lambda 2803.5, 2796.4$)
are saturated in the dominating neutral components A, E, K, O, and P. As
there are no weaker transitions available for these species, the 
column densities derived for C\,{\sc ii} and Mg\,{\sc ii} in these 
components are relatively uncertain (see Table 1). For O\,{\sc i}, N\,{\sc i},
Al\,{\sc ii}, Al\,{\sc iii}, Si\,{\sc ii}, and Fe\,{\sc ii} unsaturated
lines are available that allow us to derive accurate column
densities from profile fitting. N\,{\sc i} absorption is detected
solely in component A. The measured equivalent width of the N\,{\sc i}
feature in the $\lambda 1199.6$ line (see Fig.\,3, third panel in the left row) 
is only $1.8$ m\AA, corresponding to a $3.0\sigma$ 
significance at the measured local S/N of $\sim 130$ per resolution element.
Remarkable is the detection of P\,{\sc ii} in outermost blue
components O and P, as shown in Fig.\,4. This is 
surprising, since P\,{\sc ii} has a relatively low cosmic
abundance (log (P/H)$_{\sun}=-6.44$) and is not observed 
in the strongest neutral component A.

\begin{table*}[ht!]
\caption[]{Logarithmic ion column densities and $b$ values for the
$z=2.187$ absorber towards HE\,0001-2340}
\begin{scriptsize}
\begin{tabular}{crllllcrllll}
\hline
Comp. & $v_{\rm rel}$  & $z_{\rm abs}$ & Ion           & log $N$\,$^{\rm a}$ & $b$\,$^{\rm a}$ &
Comp. & $v_{\rm rel}$  & $z_{\rm abs}$ & Ion           & log $N$\,$^{\rm a}$ & $b$\,$^{\rm a}$ \\
      &  [km\,s$^{-1}$] &                &               &                     & [km\,s$^{-1}$] &
      &  [km\,s$^{-1}$] &                &               &                     & [km\,s$^{-1}$] \\
\hline
A & $+13$  & $2.18714$ & C\,{\sc ii}   & $14.36\pm0.19$ & $4.5\pm0.8$ & I & $-240$ & $2.18445$ & C\,{\sc ii}   & $13.03\pm0.05$ & $4.8\pm1.3$ \\
  &        &           & N\,{\sc i}    & $12.05\pm0.03$ &             &   &        &           & Mg\,{\sc ii}  & $12.23\pm0.05$ &             \\          
  &        &           & N\,{\sc ii}   & $\leq12.80$    &             &   &        &           & Al\,{\sc ii}  & $11.23\pm0.04$ &             \\
  &        &           & O\,{\sc i}    & $14.48\pm0.05$ &             &   &        &           & Si\,{\sc ii}  & $12.31\pm0.04$ &             \\
  &        &           & Mg\,{\sc ii}  & $13.13\pm0.10$ &             &   &        &           & Fe\,{\sc ii}  & $11.80\pm0.05$ &             \\
  &        &           & Al\,{\sc ii}  & $11.98\pm0.04$ &             &   &        &           &               &                &             \\
  &        &           & Al\,{\sc iii} & $11.45\pm0.04$ &             & J & $-271$ & $2.18412$ & C\,{\sc ii}   & $13.11\pm0.05$ & $5.8\pm1.1$ \\
  &        &           & Si\,{\sc ii}  & $13.35\pm0.04$ &             &   &        &           & Mg\,{\sc ii}  & $12.07\pm0.05$ &             \\
  &        &           & P\,{\sc ii}   & $\leq 12.02$   &             &   &        &           & Al\,{\sc ii}  & $11.28\pm0.05$ &             \\
  &        &           & S\,{\sc ii}   & $\leq 13.12$   &             &   &        &           & Si\,{\sc ii}  & $12.11\pm0.05$ &             \\
  &        &           & Fe\,{\sc ii}  & $12.98\pm0.03$ &             &   &        &           & Fe\,{\sc ii}  & $11.76\pm0.05$ &             \\
  &        &           & Fe\,{\sc iii} & $\leq 12.67$   &             &   &        &           &               &                &             \\
  &        &           & Zn\,{\sc ii}  & $\leq 11.04$   &             & K & $-284$ & $2.18398$ & C\,{\sc ii}   & $13.40\pm0.03$ & $4.5\pm1.0$ \\
  &        &           & H$_2$         & $\leq 13.30$   &             &   &        &           & O\,{\sc i}    & $12.85\pm0.05$ &             \\
  &        &           &               &                &             &   &        &           & Mg\,{\sc ii}  & $12.46\pm0.04$ &             \\
B & $-29$  & $2.18669$ & C\,{\sc ii}   & $13.20\pm0.04$ & $6.0\pm1.0$ &   &        &           & Al\,{\sc ii}  & $11.71\pm0.04$ &             \\
  &        &           & O\,{\sc i}    & $12.67\pm0.05$ &             &   &        &           & Al\,{\sc iii} & $11.05\pm0.03$ &             \\
  &        &           & Mg\,{\sc ii}  & $12.07\pm0.04$ &             &   &        &           & Si\,{\sc ii}  & $12.56\pm0.05$ &             \\
  &        &           & Al\,{\sc ii}  & $11.06\pm0.05$ &             &   &        &           & Fe\,{\sc ii}  & $11.99\pm0.04$ &             \\
  &        &           & Si\,{\sc ii}  & $12.32\pm0.03$ &             &   &        &           &               &                &             \\
  &        &           & Fe\,{\sc ii}  & $11.80\pm0.05$ &             & L & $-305$ & $2.18376$ & C\,{\sc ii}   & $13.17\pm0.05$ & $4.9\pm0.7$ \\
  &        &           &               &                &             &   &        &           & O\,{\sc i}    & $12.58\pm0.06$ &             \\
C & $-111$ & $2.18582$ & C\,{\sc ii}   & $12.43\pm0.03$ & $3.0\pm0.8$ &   &        &           & Mg\,{\sc ii}  & $12.17\pm0.03$ &             \\
  &        &           & Mg\,{\sc ii}  & $11.83\pm0.03$ &             &   &        &           & Al\,{\sc ii}  & $11.41\pm0.03$ &             \\
  &        &           & Al\,{\sc ii}  & $10.89\pm0.04$ &             &   &        &           & Si\,{\sc ii}  & $12.25\pm0.03$ &             \\ 
  &        &           & Si\,{\sc ii}  & $11.31\pm0.04$ &             &   &        &           & Fe\,{\sc ii}  & $11.72\pm0.03$ &             \\
  &        &           &               &                &             &   &        &           &               &                &             \\
D & $-145$ & $2.18546$ & C\,{\sc ii}   & $12.96\pm0.06$ & $5.0\pm1.1$ & M & $-327$ & $2.18352$ & C\,{\sc ii}   & $12.63\pm0.06$ & $5.0\pm1.7$ \\
  &        &           & Mg\,{\sc ii}  & $11.88\pm0.06$ &             &   &        &           & Mg\,{\sc ii}  & $11.93\pm0.05$ &             \\
  &        &           & Al\,{\sc ii}  & $11.28\pm0.03$ &             &   &        &           & Al\,{\sc ii}  & $11.49\pm0.05$ &             \\
  &        &           & Si\,{\sc ii}  & $12.16\pm0.04$ &             &   &        &           & Si\,{\sc ii}  & $11.76\pm0.05$ &             \\
  &        &           &               &                &             &   &        &           &               &                &             \\
E & $-159$ & $2.18531$ & C\,{\sc ii}   & $13.94\pm0.10$ & $3.1\pm0.9$ & N & $-340$ & $2.18339$ & C\,{\sc ii}   & $12.59\pm0.09$ & $6.4\pm2.5$ \\
  &        &           & N\,{\sc ii}   & $\leq 12.82$   &             &   &        &           & Mg\,{\sc ii}  & $11.63\pm0.07$ &             \\
  &        &           & O\,{\sc i}    & $13.13\pm0.04$ &             &   &        &           & Al\,{\sc ii}  & $11.26\pm0.06$ &             \\
  &        &           & Mg\,{\sc ii}  & $12.81\pm0.06$ &             &   &        &           & Si\,{\sc ii}  & $11.73\pm0.05$ &             \\
  &        &           & Al\,{\sc ii}  & $11.93\pm0.03$ &             &   &        &           &               &                &             \\
  &        &           & Al\,{\sc iii} & $11.09\pm0.04$ &             & O & $-353$ & $2.18325$ & C\,{\sc ii}   & $13.56\pm0.09$ & $4.3\pm1.3$ \\
  &        &           & Si\,{\sc ii}  & $12.88\pm0.03$ &             &   &        &           & O\,{\sc i}    & $12.41\pm0.05$ &             \\
  &        &           & Fe\,{\sc ii}  & $12.35\pm0.03$ &             &   &        &           & Mg\,{\sc ii}  & $12.76\pm0.05$ &             \\
  &        &           & Fe\,{\sc iii} & $\leq 12.75$   &             &   &        &           & Al\,{\sc ii}  & $12.30\pm0.05$ &             \\
  &        &           &               &                &             &   &        &           & Al\,{\sc iii} & $11.83\pm0.05$ &             \\
F & $-187$ & $2.18501$ & C\,{\sc ii}   & $13.02\pm0.05$ & $5.9\pm1.4$ &   &        &           & Si\,{\sc ii}  & $13.00\pm0.04$ &             \\
  &        &           & Mg\,{\sc ii}  & $11.46\pm0.05$ &             &   &        &           & P\,{\sc ii}   & $12.04\pm0.04$ &             \\
  &        &           &               &                &             &   &        &           & Fe\,{\sc ii}  & $11.75\pm0.04$ &             \\

G & $-217$ & $2.18469$ & C\,{\sc ii}   & $12.75\pm0.06$ & $6.6\pm2.3$ &   &        &           & Fe\,{\sc iii} & $\leq 12.79$   &             \\
  &        &           & Mg\,{\sc ii}  & $11.79\pm0.05$ &             &   &        &           &               &                &             \\
  &        &           & Al\,{\sc ii}  & $10.99\pm0.06$ &             & P & $-369$ & $2.18308$ & C\,{\sc ii}   & $13.59\pm0.07$ & $5.1\pm1.4$ \\
  &        &           & Si\,{\sc ii}  & $11.68\pm0.07$ &             &   &        &           & O\,{\sc i}  & $\leq 12.34$ &             \\ 
  &        &           &               &                &             &   &        &           & Mg\,{\sc ii}  & $12.87\pm0.04$ &             \\
H & $-229$ & $2.18457$ & C\,{\sc ii}   & $12.94\pm0.10$ & $3.5\pm1.3$ &   &        &           & Al\,{\sc ii} & $12.44\pm0.04$ &             \\
  &        &           & Mg\,{\sc ii}  & $12.01\pm0.07$ &             &   &        &           & Al\,{\sc iii}  & $12.15\pm0.04$ &             \\
  &        &           & Al\,{\sc ii}  & $11.00\pm0.05$ &             &   &        &           & Si\,{\sc ii}   & $13.08\pm0.04$ &             \\
  &        &           & Si\,{\sc ii}  & $11.95\pm0.05$ &             &   &        &           & P\,{\sc ii}  & $12.22\pm0.04$ &             \\
  &        &           & Fe\,{\sc ii}  & $11.53\pm0.04$ &             &   &        &           & Fe\,{\sc ii} & $\leq 11.82$   &             \\
& & & & & & & & &   Fe\,{\sc iii} & $\leq 12.81$   &             \\
\hline
\end{tabular}
\noindent
\\
$^{\rm a}$\, $1\sigma$ errors and $3\sigma$ upper limits are given.
\end{scriptsize}
\end{table*}

\begin{figure}
\resizebox{1.0\hsize}{!}{\includegraphics{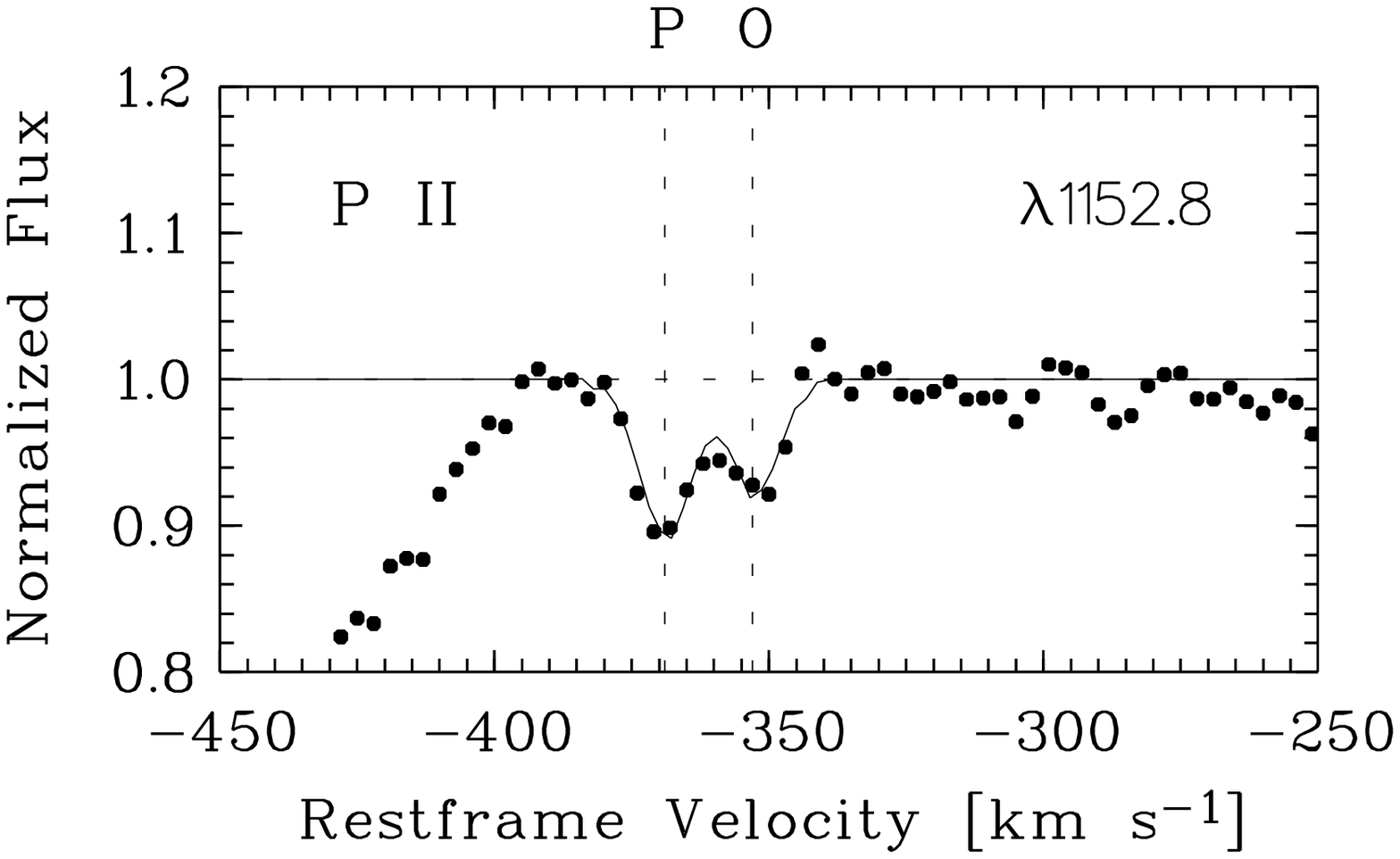}}
\caption[]{
Zoomed-in version of P\,{\sc ii} $\lambda 1152.8$
absorption in the $z=2.187$ absorber in the range
between $-450$ and $-250$ km\,s$^{-1}$. P\,{\sc ii}
absorption in components O and P (indicated by the dashed lines) is
clearly detected. The solid line shows 
the optimum two-component Voigt-profile fit.
}
\end{figure}

\subsection{Neutral hydrogen absorption}

\begin{figure*}[t!]
\resizebox{1.0\hsize}{!}{\includegraphics{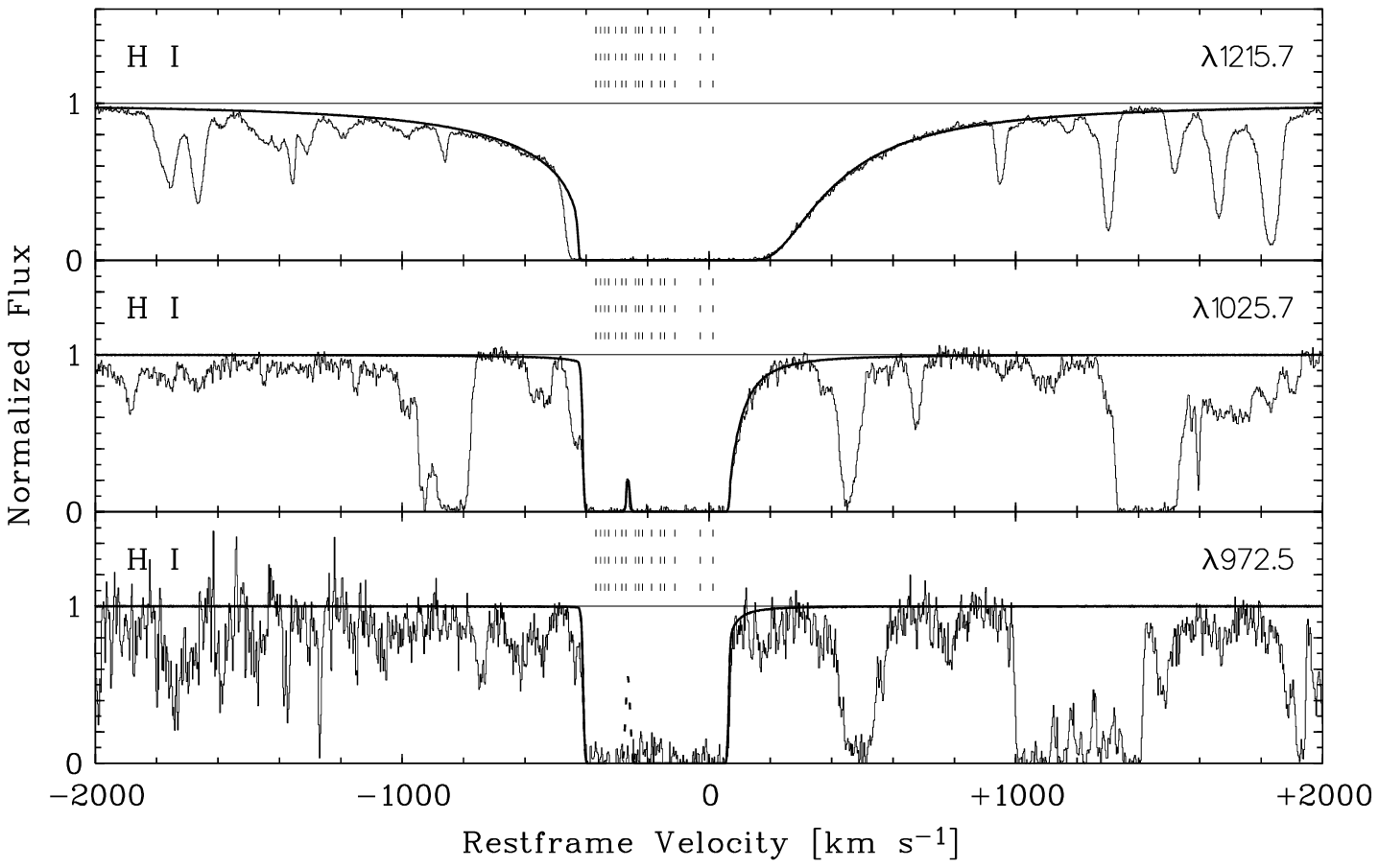}}
\caption[]{
H\,{\sc i} Ly\,$\alpha$, Ly\,$\beta$, and Ly\,$\gamma$
absorption in the $z=2.187$ absorber. The thick solid line
represents the best-fitting H\,{\sc i} model, as 
described in Sect.\,3.3.
}
\end{figure*}

Neutral hydrogen absorption in the $z=2.187$ sub-DLA is 
seen in Ly\,$\alpha$, Ly\,$\beta$, and Ly\,$\gamma$, 
as shown in Fig.\,5. The 
other lines from the H\,{\sc i} Lyman series either have 
very low S/N or are located
bluewards of the available UVES wavelength range.
As clearly visible in Figs.\,1 and 5, the Ly\,$\alpha$
absorption is very asymmetric due to the large number
of neutral gas components at the blue side that contribute to the
overall H\,{\sc i} absorption. The red part of the
Ly\,$\alpha$ absorption shows a well-defined damping
wing caused by the strongest neutral gas component 
at $+13$ km\,s$^{-1}$ (component A).
Due to the asymmetry, a free single-component fit of
the Ly\,$\alpha$, Ly\,$\beta$, and Ly\,$\gamma$ absorption
cannot even roughly reproduce the observed shape of the absorption.
Therefore, the complex velocity structure of the $z=2.187$ 
absorber has to be taken into account for the 
profile fit of the H\,{\sc i} Lyman series absorption.
Fixing the centroid of the fit at $+13$ km\,s$^{-1}$ 
and only fitting the damping wing on the red side of the 
Ly\,$\alpha$, Ly\,$\beta$, and Ly\,$\gamma$ absorption simultaneously results
in an H\,{\sc i} column density of 
log $N$(H\,{\sc i}$)_{\rm A}=19.60\pm0.4$
for component A. Similarly, we can constrain the total H\,{\sc i}
column density in the two bluest velocity components
O and P to log $N$(H\,{\sc i}$)_{\rm OP}\leq 18$ by
fixing the velocity centroid at $-361$ km\,s$^{-1}$ and
fitting the blue wing of the Ly\,$\beta$ and Ly\,$\gamma$ absorption.
Note that the additional absorption
occurring in the blue wing of the Ly\,$\alpha$, Ly\,$\beta$,
and Ly\,$\gamma$ absorption is possibly
due to deuterium (D\,{\sc i}) in
components O and P in combination with
an additional weak H\,{\sc i} component near
$-440$ km\,s$^{-1}$. This will be further
discussed in the Appendix.

The entire H\,{\sc i} absorption in Ly\,$\alpha$, Ly\,$\beta$, 
and Ly\,$\gamma$ profile can be modeled
by taking into account the absorption from the intermediate
velocity components. For our model, we consider absorption
from the strongest neutral gas components A, B, E, K,
L, O, and P, for which we have fixed the velocity-component
structure as seen in the metal-line absorption.
Characteristic is the presence of a 
residual flux spike in the Ly\,$\beta$ profile near
$-300$ km\,s$^{-1}$, which needs to be reproduced by
our model. A similar spike is not seen in the Ly\,$\gamma$
absorption profile, but it is likely that in this region the Ly\,$\gamma$
line is blended with absorption from the Ly\,$\alpha$ forest.
The best fitting model has logarithmic H\,{\sc i} column densities of $19.60$, $17.79$,
$18.25$, $15.03$, $17.71$, $17.53$, and $17.32$ for the
components A, B, E, K, L, O, \& P, respectively.
This model is shown Fig.\,5 with a thick solid line;
the expected flux spike near $-300$ km\,s$^{-1}$ 
in the Ly\,$\gamma$ absorption is indicated with
a thick dashed line.
Note that this fit for the intermediate velocity components is 
not unique, as other combinations of H\,{\sc i} column densities 
also can reproduce the shape of the observed H\,{\sc i} absorption.
An important outcome of this model is, however, that
the absorption of the intermediate components 
does not significantly
influence the shape of the H\,{\sc i} absorption in blue and red wing
of the Lyman lines. Therefore, the values of
log $N$(H\,{\sc i}$)_{\rm A}=19.60\pm0.4$
and log $N$(H\,{\sc i}$)_{\rm OP}\leq 18$ derived above represent
reliable estimates for the H\,{\sc i} column densities
in components A and O$+$P.

\begin{table}[t]
\caption[]{Summary of logarithmic metal abundances 
in component A}
\begin{normalsize}
\begin{tabular}{llll}
\hline
Species & I.P. & log (X/H)$_{\sun}$\,$^{\rm a}$ & [X/H]$^{\rm b}$\\
        & [eV] & (+12)                          & \\
\hline
C\,{\sc ii}  & 24.38 & 8.52 & $-1.76\pm0.21$ \\
N\,{\sc i}   & 14.53 & 7.95 & $-3.50\pm0.05$ \\
N\,{\sc ii}  & 29.60 & 7.95 & $\leq -2.75$ \\
O\,{\sc i}   & 13.62 & 8.69 & $-1.81\pm0.07$ \\
Mg\,{\sc ii} & 15.04 & 7.58 & $-2.05\pm0.11$ \\
Al\,{\sc ii} & 18.83 & 6.49 & $-2.11\pm0.06$ \\
Al\,{\sc iii}& 28.45 & 6.49 & $-2.64\pm0.06$ \\
Si\,{\sc ii} & 16.35 & 7.56 & $-1.81\pm0.07$ \\
P\,{\sc ii}  & 19.77 & 5.56 & $\leq -1.18$ \\
S\,{\sc ii}  & 23.34 & 7.20 & $\leq -1.68$ \\
Fe\,{\sc ii} & 16.19 & 7.50 & $-2.12\pm0.05$ \\
Fe\,{\sc iii} & 30.65 & 7.50 & $\leq -2.23$ \\
Zn\,{\sc ii} & 17.96 & 4.67 & $\leq -1.23$ \\
\hline
\end{tabular}
\noindent
\\
{\small 
$^{\rm a}$\, Reference abundances are taken from Allende Prieto et al.\,(2001) for oxygen and from
   the compilation of Morton (2003) and references therein.\\
$^{\rm b}$\, [X/H]$=$log (X/H)$-$log (X/H)$_{\sun}$.}
\end{normalsize}
\end{table}

\subsection{Abundances and abundance ratios}

\subsubsection{Component A}

First, we concentrate on the dominating neutral
component (component A) in our 
discussion of abundances and abundance ratios in the 
$z=2.187$ absorber. The best
element to derive the overall abundance in the gas 
is oxygen, since neutral oxygen and neutral
hydrogen have very similar ionization potentials
and both elements are coupled by a strong charge-exchange
reaction. Moreover, oxygen does not deplete significantly
into dust grains. Our analysis
yields [O/H$]=-1.81\pm 0.07$
\footnote{[X/H]$=$log (X/H)$-$log (X/H)$_{\sun}$},
which corresponds to an abundance of $\sim 1/65$ solar. 
In striking contrast, the nitrogen abundance
is significantly lower, [N/H$]=-3.50\pm 0.05$  
or $\sim 1/3200$ solar, as derived from
N\,{\sc i}. The resulting
nitrogen-to-oxygen ratio [N/O$]=-1.69$, if accurate, would be
one of the lowest ever measured in DLAs so far.
However, the interpretation of N\,{\sc i}/H\,{\sc i} is
not as straightforward as for O\,{\sc i}/H\,{\sc i}.
Although the ionization potential of N\,{\sc i}
is only slightly higher than that of hydrogen ($14.53$ versus
$13.60$ eV),
ionization effects can significantly affect the determination of
[N/H] from N\,{\sc i} and H\,{\sc i} in absorbers with
overall column densities $\leq 10^{20}$ cm$^{-2}$. 
The reason for this is that 
charge exchange reactions are less efficient between N\,{\sc i} and
H\,{\sc i} (in contrast to O\,{\sc i} and
H\,{\sc i}; see Butler \& Dalgarno 1979) 
and N\,{\sc i} has a relatively large
photoionization cross section (Sofia \& Jenkins 1998).
Thus, a considerable fraction of the nitrogen in component
A may reside
in the N$^+$ phase, particularly if a substantial ionizing radiation 
field is present.
While the uncertain column density constraint from the blended 
N\,{\sc ii} $\lambda 1084.0$ line (log $N$(N\,{\sc ii}$)\leq 12.80$) 
does not provide a very stringent estimate of
the abundance of singly ionized nitrogen, we can 
check upon possible ionization effects in component A
using other weakly ionized species.
The ratio of (Al\,{\sc iii}/A\,{\sc ii}$)\approx0.3$ suggests
that ionization could be important. This ratio, however, is
difficult to interpret due to the uncertain recombination
rate of Al$^+$ (see Vladilo et al.\,2001; Nussbaumer \& Storey 1986)
and the observed discrepancies between (Al\,{\sc iii}/A\,{\sc ii}) and
other ionic ratios in high-z photo-ionization models 
(e.g., D'Odorico \& Petitjean 2001).
An important element to investigate ionization conditions
in interstellar and intergalactic gas is the 
$\alpha$ element sulfur
(Richter et al.\,2001). 
S\,{\sc ii} has an ionization potential of $23.23$ eV
and depletes only very little into dust, so that we
expect most S to be in form of S\,{\sc ii}. Since 
the ionization potential of S\,{\sc ii} is significantly 
higher than that of O\,{\sc i} and H\,{\sc i} 
($\sim 13.6$ eV for both), S\,{\sc ii} lives in
gas where part of the hydrogen is ionized.
From the observed S\,{\sc ii}/O\,{\sc i} ratio we thus
can learn about the hydrogen ionization fraction
(see Richter et al.\,2001 for details). Although S\,{\sc ii}
is not detected in our sub-DLA, the $3\sigma$ upper limit for the
S\,{\sc ii} column density in component A 
(log $N$(S\,{\sc ii}$)\leq13.12$) yields [S/H]$\leq -1.68$ and
[S\,{\sc ii}/O\,{\sc i}]$\leq +0.13$. If one assumes a solar sulfur-to-oxygen ratio,
the possible small excess of S\,{\sc ii} compared to O\,{\sc i} 
implies that the degree of ionization in component A is
low ($<30$ percent).
\footnote{Note that if the intrinsic S/O ratio in this system
is significantly lower than solar, the ionization fraction would be 
underestimated by this method. Some DLAs indeed show
sub-solar S/O ratios (e.g., 
Prochaska, Howk, \& Wolfe (2003)).}
We conclude that N\,{\sc ii}/N\,{\sc i} most likely is
small, so that we can give a conservative limit for
the nitrogen abundance in component A of
[N/H]$\lesssim -3.3$ and [N/O]$\lesssim -1.5$. 

The apparently low degree of
ionization in combination with 
the equal abundances of Si and O speaks against
any significant dust depletion of silicon 
in component A.
The situation is somewhat less clear for the elements
Mg, Al, and Fe, which are all underabundant compared
to oxygen ([Mg/O]$=-0.24$, [Al/O]$=-0.30$, and 
[Fe/O]$=-0.31$). These elements could be depleted into dust
grains. In fact, with [Fe/O]$=-0.31$ and [Si/O]$=0.00$ the
absorber follows the trend for dust depletion patterns
found in damped Ly\,$\alpha$ systems and in the 
Galactic halo (see Petitjean, Srianand, \& Ledoux 2002,
their Fig.\,7). However, next to dust depletion 
nucleosynthesis effects certainly have to be considered
for these elements given the low overall abundance and
the very low nitrogen content. In this context, the
observed abundances of iron and aluminum could also
be explained by a slightly enhanced [$\alpha$/Fe] element
ratio and a mild odd-even effect for Al. In contrast
to Fe and Al,
the column density of magnesium 
unfortunately is too uncertain (due 
to saturation of the two available Mg\,{\sc ii} 
lines) to draw meaningful conclusions about the
apparent Mg underabundance.
The absence of molecular hydrogen absorption in
component A represents yet another piece of 
information on the dust abundance in the gas.
From the H$_2$ column density limit (see Table 1)
we obtain an upper limit for the molecular hydrogen
fraction of $f=[2N($H$_2)/(N$(H\,{\sc i}$)+2N($H$_2))]
\approx5\times10^{-7}$.
As H$_2$ preferentially forms on the surfaces of dust grains,
the lack of H$_2$
further supports the above conclusion that
the dust abundance in component A 
must be small (see also Ledoux, Petitjean, \& 
Srianand 2003).
Little can
be said about the carbon abundance in component A, as the 
value of [C/H$]=-1.76\pm 0.21$ - although similar 
to that of oxygen - is very uncertain
due to saturation of the two C\,{\sc ii} lines.
If the $b$ value would be only slightly higher than the 
assumed $4.5$ km\,s$^{-1}$, the C\,{\sc ii}
column density (and thus the derived carbon abundance) would
drop substantially. 

All measured abundances in component A are 
summarized in Table 2.

\subsubsection{Components O and P}

We now turn to components O and P, which
show a very distinct abundance pattern.
As discussed in Sect.\,3.3, the Ly\,$\alpha$ 
Voigt profile fit yields log $N$(H\,{\sc i}$)
\leq 18$ for components O and P together.
The combined O\,{\sc i} column
density for O and P is log $N \approx 12.7$ (see
Table 1), from which we infer an oxygen abundance
of [O/H]$_{\rm OP}]\geq -2.0$. 
This is consistent with the oxygen abundance derived 
for component A.
The large value for log (C\,{\sc ii}/O\,{\sc i})$=+1.15$ 
in component O implies that the degree of ionization
must be large. Note that the C\,{\sc ii} absorption
in components O and P is not saturated (unlike
in component A), so that the derived 
C\,{\sc ii} column densities are 
accurate (see Table 1).
From C\,{\sc ii}, Mg\,{\sc ii}, 
Si\,{\sc ii}, Al\,{\sc ii},
and P\,{\sc ii} we obtain
(Mg\,{\sc ii}/C\,{\sc ii}$)_{\rm OP}=-0.76$,
(Si\,{\sc ii}/C\,{\sc ii}$)_{\rm OP}=-0.54$,
(Al\,{\sc ii}/C\,{\sc ii}$)_{\rm OP}=-1.21$,
and (P\,{\sc ii}/C\,{\sc ii}$)_{\rm OP}=-1.45$.
From Fe\,{\sc ii} and Fe\,{\sc iii}
we derive ((Fe\,{\sc ii}$+$Fe\,{\sc iii})/C\,{\sc ii}$)_{\rm OP}=-0.74$.
These ratios suggest significant overabundances
of silicon, aluminum, and phosphorus in
components O and P.
However, a confirmation of this 
statement and a more quantitative estimate
of the apparent overabundances 
has to await a detailed photoionization
modeling for these species. 
This will be presented in Sect.\,4.2.

\section{Highly-ionized species}

\subsection{Component structure and column densities}

Absorption by highly ionized gas in the $z=2.187$ absorber
is seen in C\,{\sc iv} $\lambda \lambda 1550.8, 1548.2$ and
Si\,{\sc iv} $\lambda \lambda 1402.8, 1393.3$. O\,{\sc vi} and
N\,{\sc v} are not detected, but blending problems hamper
a reliable detection of these ions. The component structure seen in
C\,{\sc iv} and Si\,{\sc iv} (Fig.\,6) is even more complex than that of
the neutral and weakly ionized gas. All in all, we identify 
25 absorption components including the 16 components detected
in neutral and weakly ionized species. All C\,{\sc iv} and 
Si\,{\sc iv} measurements are
summarized in Table 3.

\begin{figure}[t!]
\resizebox{1.0\hsize}{!}{\includegraphics{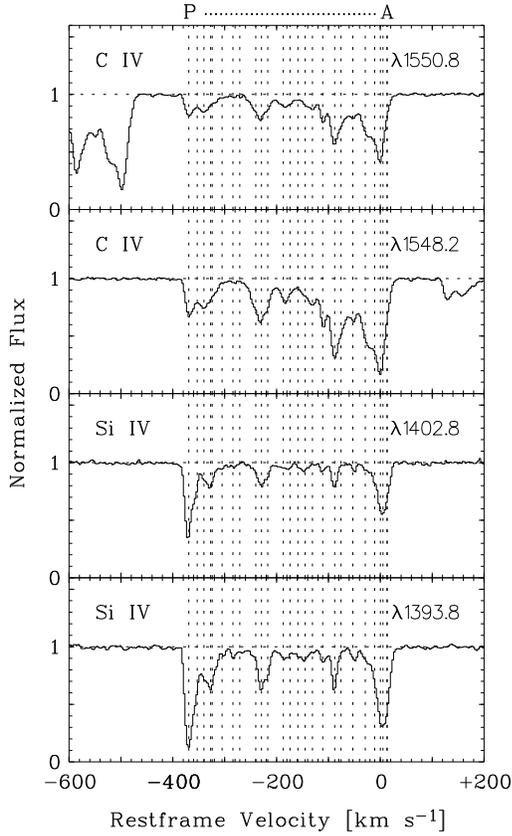}}
\caption[]{
Absorption by C\,{\sc iv} ($\lambda\lambda 1550.8, 1548.2$)
and Si\,{\sc iv} ($\lambda\lambda 1402.8, 1393.8$) in the
$z=2.187$ absorber is shown. The absorption pattern of these high ions
is even more complex than for the neutral and weakly ionized
species. We identify 25 components that partly correspond
to the absorption components seen in C\,{\sc ii} (see
also Table 3 for details).
}
\end{figure}

\begin{table}[t!]
\caption[]{Logarithmic column densities for C\,{\sc iv} and Si\,{\sc iv}}
\begin{scriptsize}
\begin{tabular}{lrlllr}
\hline
Comp. & $v_{\rm rel}$ & $z_{\rm abs}$ & Ion & log $N$\,$^{\rm a}$ & $b$\,$^{\rm a}$ \\
      & [km\,s$^{-1}$] &                &     &                     & [km\,s$^{-1}$] \\
\hline
A   &  $+13$ & $2.18714$ & C\,{\sc iv}  & $12.01\pm0.05$ & $4.5\pm0.7$ \\
    &        &           & Si\,{\sc iv} & $12.28\pm0.04$ & $5.6\pm0.4$ \\
\\
A1  &   $+5$ & $2.18705$ & Si\,{\sc iv} & $13.01\pm0.03$ & $9.4\pm1.1$ \\
\\
A2  &    $0$ & $2.18700$ & C\,{\sc iv}  & $13.58\pm0.04$ & $9.7\pm1.4$ \\
\\
A3  &  $-11$ & $2.18688$ & Si\,{\sc iv} & $12.52\pm0.08$ & $18.5\pm3.6$ \\
\\
B   &  $-29$ & $2.19669$ & C\,{\sc iv}  & $13.48\pm0.06$ & $14.6\pm2.7$ \\
\\
B1  &  $-53$ & $2.18644$ & C\,{\sc iv}  & $12.84\pm0.11$ & $11.6\pm4.4$ \\
    &        &           & Si\,{\sc iv} & $12.11\pm0.14$ & $3.5\pm1.8$ \\
\\
B2  &  $-76$ & $2.18619$ & C\,{\sc iv}  & $13.02\pm0.12$ & $8.8\pm4.9$ \\
\\
B3  &  $-88$ & $2.18606$ & C\,{\sc iv}  & $13.22\pm0.07$ & $6.8\pm0.9$ \\
    &        &           & Si\,{\sc iv} & $12.45\pm0.04$ & $4.7\pm0.3$ \\
\\
C   & $-111$ & $2.18582$ & C\,{\sc iv}  & $12.93\pm0.06$ & $6.5\pm0.8$ \\
    &        &           & Si\,{\sc iv} & $11.85\pm0.05$ & $4.9\pm1.3$ \\
\\
C1  & $-131$ & $2.18561$ & C\,{\sc iv}  & $12.63\pm0.06$ & $7.6\pm1.1$ \\
\\
D   & $-145$ & $2.18546$ & C\,{\sc iv}  & $12.40\pm0.11$ & $6.5\pm2.4$ \\
    &        &           & Si\,{\sc iv} & $11.74\pm0.06$ & $6.4\pm1.8$ \\
\\
E   & $-159$ & $2.18531$ & Si\,{\sc iv} & $11.86\pm0.23$ & $15.4\pm4.4$\\
\\
E1  & $-174$ & $2.18515$ & C\,{\sc iv}  & $11.65\pm0.07$ & $9.7\pm2.0$ \\
\\
F   & $-187$ & $2.18501$ & C\,{\sc iv}  & $12.72\pm0.10$ & $11.2\pm2.1$ \\
    &        &           & Si\,{\sc iv} & $11.90\pm0.06$ & $11.4\pm1.7$ \\
\\
G   & $-217$ & $2.18469$ & C\,{\sc iv}  & $12.75\pm0.04$ & $8.8\pm1.8$ \\
    &        &           & Si\,{\sc iv} & $12.06\pm0.05$ & $3.1\pm1.1$ \\
\\
H   & $-229$ & $2.18457$ & C\,{\sc iv}  & $12.90\pm0.04$ & $9.0\pm1.3$ \\
    &        &           & Si\,{\sc iv} & $12.49\pm0.03$ & $6.6\pm0.5$ \\
\\
I   & $-240$ & $2.18445$ & C\,{\sc iv}  & $12.58\pm0.08$ & $9.4\pm2.2$\\
    &        &           & Si\,{\sc iv} & $11.80\pm0.09$ & $9.3\pm1.5$ \\
\\
J   & $-271$ & $2.18412$ & Si\,{\sc iv} & $11.68\pm0.13$ & $9.8\pm3.4$ \\
\\
K   & $-284$ & $2.18398$ & Si\,{\sc iv} & $11.55\pm0.17$ & $2.2\pm1.6$ \\
\\
L   & $-305$ & $2.18376$ & Si\,{\sc iv} & $11.80\pm0.10$ & $5.0\pm2.8$\\
\\
L1  & $-324$ & $2.18356$ & Si\,{\sc iv} & $12.11\pm0.08$ & $4.1\pm1.7$\\
\\
M   & $-327$ & $2.18352$ & C\,{\sc iv}  & $12.63\pm0.05$ & $14.2\pm3.0$ \\
    &        &           & Si\,{\sc iv} & $12.40\pm0.07$ & $9.0\pm1.3$ \\
\\
N   & $-340$ & $2.18339$ & C\,{\sc iv}  & $12.74\pm0.05$ & $9.5\pm2.3$ \\
    &        &           & Si\,{\sc iv} & $12.09\pm0.04$ & $2.0\pm0.3$ \\
\\
O   & $-353$ & $2.18325$ & C\,{\sc iv}  & $12.41\pm0.07$ & $8.9\pm2.4$ \\
    &        &           & Si\,{\sc iv} & $12.69\pm0.05$ & $7.2\pm2.2$ \\
\\
P   & $-369$ & $2.18308$ & C\,{\sc iv}  & $12.88\pm0.06$ & $8.5\pm1.6$ \\
    &        &           & Si\,{\sc iv} & $13.20\pm0.04$ & $5.5\pm1.0$ \\
\hline
\end{tabular}
\noindent
\\
$^{\rm a}$\, $1\sigma$ errors and $3\sigma$ upper limits are given.
\end{scriptsize}
\end{table}

\begin{figure}[t!]
\resizebox{0.95\hsize}{!}{\includegraphics{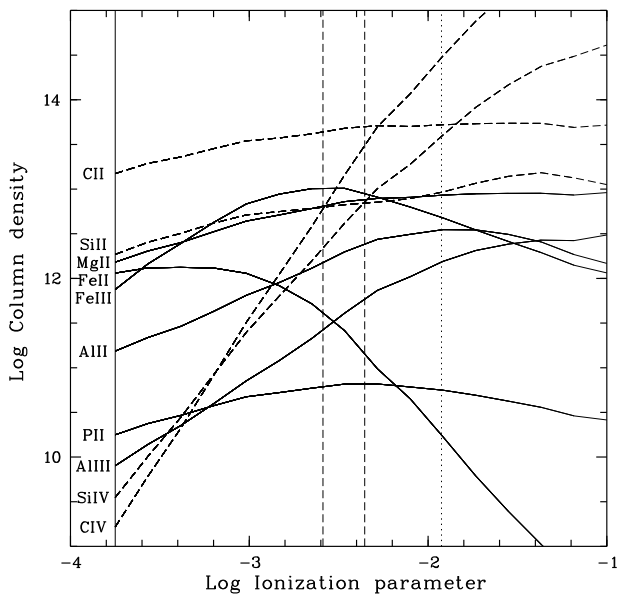}}
\caption[]{
CLOUDY photoionization model for several ions detected in
components O and P. The model assumes an overall abundance
of $0.01$ solar together with a Madau, Haardt, \& Rees (1999)
metagalactic ionizing spectrum for $z=2.2$. 
The dashed vertical lines indicate the fits for the 
C\,{\sc iv}/C\,{\sc ii} (left) and Si\,{\sc iv}/Si\,{\sc ii} (right)
column density ratios, the dotted vertical line 
indicates the fit for Al\,{\sc iii}/Al\,{\sc ii} 
(see Sect.\,4.2
for details).
}
\end{figure}

\subsection{Photoionization modeling for comp. O and P}

Our measurements of C\,{\sc ii}, C\,{\sc iv}, Si\,{\sc ii}, 
and Si\,{\sc iv} allow us
to model the photoionization in components O and P
and to constrain the ionization parameter
$U=n_{\gamma}/n_{\rm H}$,
the ratio of ionizing photon density to gas density.
We have used the photoionization code
CLOUDY v94.00 (Ferland 1996) and 
have modeled a plane-parallel slab of gas with a metallicity
of 0.01 solar and solar relative chemical
abundances (Fig.\,7). The calculations were done using a constant
density and were stopped whenever the neutral hydrogen
column density reached $10^{18}$ cm$^{-2}$.
The total hydrogen
column density in the model, $N$(H\,{\sc i}+H\,{\sc ii}),
comes out to $1.0 \times 10^{20}$ cm$^{-2}$.
For the ionizing radiation, we used the
Madau, Haardt \& Rees (1999) metagalactic spectrum at
$z=2.2$.
Our CLOUDY modeling for the column density ratios of
C\,{\sc iv}/C\,{\sc ii} and Si\,{\sc iv}/Si\,{\sc ii}
implies ionization parameters of log $U=-2.59$ and
$-2.35$, respectively, thus in good agreement with
each other (see Fig.\,7, vertical dashed lines). 
A value of log $U\approx-2.5$ further is consistent 
with the observed column densities/column density limits
for Fe\,{\sc ii} and Fe\,{\sc iii}.
From the Al\,{\sc iii}/Al\,{\sc ii} ratio we
obtain a significantly higher value of log $U=-1.92$
(Fig.\,7, vertical dotted line). This high value
most likely is a result of the poorly known
dielectric recombination rate of Al$^+$ at high temperatures
(Petitjean, Rauch, \& Carswell 1994; Nussbaumer \&
Storey 1986) and therefore will not be considered here
any further. Comparing the column densities predicted
by the CLOUDY model with those measured allows us 
to constrain the metal abundances in components O and P.
The results are summarized in Table 4. For oxygen, carbon,
magnesium, and iron we
obtain abundances that are consistent with those 
measured in component A. For aluminum, silicon, and
in particular for phosphorus, the CLOUDY model results
in relative abundances that are significantly enhanced
compared to solar. 
The overabundances relative to carbon amount to $0.5\pm 0.1$ dex 
for aluminum, $0.6\pm 0.1$ dex for silicon, and 
$1.5\pm 0.1$ dex for phosphorus (see Table 4).

\begin{table}[t!]
\caption[]{Summary of logarithmic metal abundances
in components O \& P}
\begin{normalsize}
\begin{tabular}{llrr}
\hline
Species &  log (X/H)$_{\sun}$\,$^{\rm a}$ & [X/H]\,$^{\rm b}$ & [X/C]\,$^{\rm b}$ \\
        & (+12)      &                  & \\
\hline
C  & 8.52 & $-1.9\pm 0.1$ & ... \\
O  & 8.69 & $-2.0\pm 0.1$ & $-0.1\pm 0.1$ \\
Mg & 7.58 & $-1.7\pm 0.1$ & $+0.2\pm 0.1$ \\
Al & 6.49 & $-1.4\pm 0.1$ & $+0.5\pm 0.2$ \\
Si & 7.56 & $-1.3\pm 0.1$ & $+0.6\pm 0.2$ \\
P  & 5.56 & $-0.4\pm 0.1$ & $+1.5\pm 0.1$ \\
Fe & 7.50 & $\leq -1.9$ & $\leq 0.0$ \\
\hline
\end{tabular}
\noindent
\\
{\small
$^{\rm a}$\, Reference abundances are taken from Allende Prieto et al.\,(2001) for oxygen and from
the compilation of Morton (2003) and references therein.\\
$^{\rm b}$\, Based on the CLOUDY photoionization model
described in Sect.\,4.2. }
\end{normalsize}
\end{table}

\section{Discussion}

Our measurements of the sub-DLA at $z\approx2.2$ towards HE\,0001$-$2340
unveil a number of interesting abundance properties in this 
absorption system. 
With [O/H]$=-1.81\pm0.07$, [N/H]$\lesssim -3.3$, and [N/O]$\lesssim -1.5$
the sub-DLA towards HE\,0001$-$2340 exhibits a classic massive star abundance
pattern with little sign of primary nitrogen (e.g., Umeda \& Nomoto
2004). While the alpha-element abundances [O/H] and [Si/H]
are comparable to those measured for other
sub-DLAs and DLAs at similar 
redshifts (e.g., Dessauges-Zavadsky et al.\,2002;
Pettini et al.\,2002), the [N/$\alpha$] ([N/O]) ratio is among the 
lowest ever measured in DLAs and other astrophysical
sites. [N/$\alpha$] is
a particularly important parameter
to trace the chemical evolution in the early Universe.
Nitrogen is believed to be produced mainly in 
intermediate-mass stars. 
The nitrogen deficiency observed in the intergalactic
medium and in blue compact dwarf (BCD) galaxies 
(e.g., Prochaska et al.\,2002; Izotov \& Thuan 1999)
probably is a result from the early chemical enrichment of
Type II supernovae from
massive stars that produce only very little
primary nitrogen. In low-metallicity environments
with [$\alpha$/H$]\leq-1.5$,
a first [N/$\alpha$] plateau possibly is present near $-1.5$ dex,
whereas for more metal-rich systems a second plateau
occurs near $-0.9$ dex (Centuri\a'on et al.\,2003).
The region around [N/$\alpha]=-0.9$ is much more
populated than the region near $-1.5$ dex,
showing that most of observed environments must 
have experienced a second epoch of 
nitrogen and $\alpha$-element enrichment
after the initial enrichment
to the [N/$\alpha]=-1.5$ level.
In contrast to the $-0.9$ dex plateau, 
that near $-1.5$ dex exhibits only small
scatter ($\sim 0.05$ dex) in (N/O).
To our knowledge, no systems with [N/O] ratios $\leq -1.70$
have been found so far.
This may suggest that the value of
[N/O]$\approx-1.5$ represents the ground floor
for (N/O) in the Universe, set by the
coeval injection of primary nitrogen and oxygen
from the first generation of (massive) stars.
The results presented here support this scenario.
However, 
the number of systems with very low (N/O) values
is still very limited and thus
the existence of the $-1.5$ dex plateau
still has to be confirmed by additional
measurements
(Centuri\a'on et al.\,2003; Israelian et al.\,2004).

Another interesting result of our study is the 
abnormal abundance pattern seen in components O and P.
The ion column densities in these components
derived in Sect.\,3.4.2 and the CLOUDY model presented
in Sect.\,4.2 imply significant overabundances of
aluminum ($0.5$ dex), silicon ($0.6$ dex),
and phosphorus ($1.5$ dex) compared to carbon.
Such an abundance pattern is highly unusual for DLAs.
The findings may indicate that the line of sight towards
HE\,$0001-2340$ passes through a confined gaseous region 
in the DLA host that is locally polluted
with heavy elements by nearby supernova explosions.
Since the ejecta of a single supernova
do not contain enough material
to significantly enhance the heavy metal abundance
in larger volumes of interstellar gas,
it seems plausible that the chemically enriched 
gas components O and P is related to 
massive (coeval) star-formation activity,
such as in massive star clusters.
Recent models for the
so-called ``hypernovae'' (supernovae that have
explosion energies $\geq10^{52}$ erg) predict
significantly enhanced [Si/C] abundances in
the supervova ejecta as a result
of carbon and oxygen burning in low-density regions
that lead to an enhanced freeze-out of $\alpha$ elements
(e.g., Nakamura et al.\,2001).
Yet, these models cannot explain overabundances
of phosphorus and aluminum and thus the exact
explosion mechanism that may lead to the observed
abundance pattern in components O and P remains unknown.

A striking feature of the $z=2.187$ absorber
towards HE\,0001$-$2340 is the large number
of individual velocity components that span a very large velocity range of 
almost $\sim 400$ km\,s$^{-1}$. This velocity span seems very large
for a single (proto)galactic structure. One possible
explanation for this is that the absorption pattern
reflects a merger event of two galaxies or protogalactic
structures. The individual velocity components
that are quite narrow but that do not contain large 
columns of gas then may arise in tidal gaseous features
produced by the merger event, spread out over 
a large spatial and radial-velocity range.
An interesting aspect of a merger scenario 
lies in the idea that the merger actually 
{\it triggers} the formation of the stellar clusters
(e.g., a nuclear cluster)
that give rise to the abundance anomalies 
in components O and P. It is well known that merger/accretion
events are of prime importance for starburst activity and the formation
of massive stellar clusters in and around galaxies 
(e.g., Knierman et al.\,2003; Schweizer \& 
Seitzer 1998). If the $z=2.187$ absorber
towards HE\,0001$-$2340 currently undergoes
a merger event that leads to the formation of
massive stellar clusters, these clusters 
should locally enrich their
interstellar environment with heavy elements by
supernova explosions.
Such circum-cluster environments, when they are
young, should not have had enough time to mix
with the more extended interstellar environment
of the host galaxy. This possibly creates distinct
regions with enhanced heavy-element abundances
similar to what is observed in components
O and P. The sub-DLA towards 
HE\,0001$-$2340 thus may represent 
a young galaxy at $z\approx2$ that 
currently forms due to merging.

\section{Summary}

We have performed a detailed analysis of the sub-DLA at $z=2.187$ 
towards HE\,0001$-$2340 and have derived the following results:\\
\\
\noindent
(1) We identify 16 absorption subcomponents in neutral and weakly ionized
species (H\,{\sc i}, D\,{\sc i}, C\,{\sc ii}, N\,{\sc i}, N\,{\sc ii}, O\,{\sc i},
Mg\,{\sc ii}, Al\,{\sc ii}, Al\,{\sc iii}, Si\,{\sc ii}, P\,{\sc ii}, and Fe\,{\sc ii}),
and 25 subcomponents in C\,{\sc iv} and Si\,{\sc iv}.
These components span a velocity range of $\sim 400$ km\,s$^{-1}$ in
the $z=2.187$ restframe. The system has a
total neutral hydrogen column density of log $N$(H\,{\sc i}$)\approx19.7$.\\
\\
\noindent
(2) The overall abundance of the absorber is [M/H]$=-1.81\pm0.07$ dex 
(1/65 solar), as derived from unsaturated O\,{\sc i} absorption in the strongest
neutral gas component A at $v_{\rm rel}=+13$ km\,s$^{-1}$
(relative to $z=2.187$). The lack of depletion of Si\,{\sc ii} and the
absence of molecular hydrogen imply that the absorber is devoid of dust.\\
\\
\noindent
(3) The gas in component A shows a classic massive star abundance
pattern with little sign of primary nitrogen. Taking
possible ionization effects into account, we estimate
[N/H]$\lesssim -3.3$ and [N/O]$\lesssim -1.5$. These
values place the $z=2.187$ absorber among systems
with the lowest measured [N/$\alpha$] ratios in
astrophysical sites so far. This result is in line
with the idea that the first stars in the Universe
have enriched the intergalactic medium to a level
of [N/$\alpha]\approx-1.5$.\\
\\
\noindent
(4) Components O and P near $-350$ km\,s$^{-1}$  
exhibit significant overabundances
of phosphorus, silicon, and aluminum 
([P/C]$\approx+1.5$, [Si/C]$\approx+0.6$, and [Al/C]$\approx+0.5$)
after applying an ionization correction.
Possibly, the gas belonging to these components
is locally enriched by the supernova ejecta from 
one or more massive stellar clusters. Absorption
by D\,{\sc i} is possibly detected in components O and P
in the blue wings of the H\,{\sc i} Ly\,$\alpha$, Ly\,$\beta$ and 
Ly\,$\gamma$ lines. However, our analysis of these lines 
leads to an extraordinarily high D/H
ratio of $\geq 1.22 \times 10^{-4}$, implying that
the D\,{\sc i} absorption is blended with an H\,{\sc i} interloper.\\
\\
\noindent
(5) We suggest that the large number of absorption
components spread over several hundred km\,s$^{-1}$ 
indicates the merging of two galaxies or protogalactic
structures. This merging eventually triggers
the formation of star clusters that are responsible
for the abundance anomalies observed in 
components O and P.
If so, the sub-DLA towards HE\,0001$-$2340 may
represent a young galaxy at $z\approx2$ that
currently forms out of a merger event.

\begin{acknowledgements}

   This work is based on observations collected during program
   166.A-0106 (PI: Jacqueline Bergeron) of the European Southern
   Observatory with the Ultraviolet and Visible Echelle
   Spectrograph mounted on the 8.2 m KUEYEN telescope operated at
   the Paranal Observatory, Chile. We are grateful to the 
   astronomers that have performed the observations in service mode.
   P.R. is supported by the German
   \emph{Deut\-sche For\-schungs\-ge\-mein\-schaft}, DFG, 
   through Emmy-Noether grant Ri 1124/3-1.

\end{acknowledgements}

{}

\appendix

\section{Comments on Deuterium}

\begin{figure}[t!]
\resizebox{1.0\hsize}{!}{\includegraphics{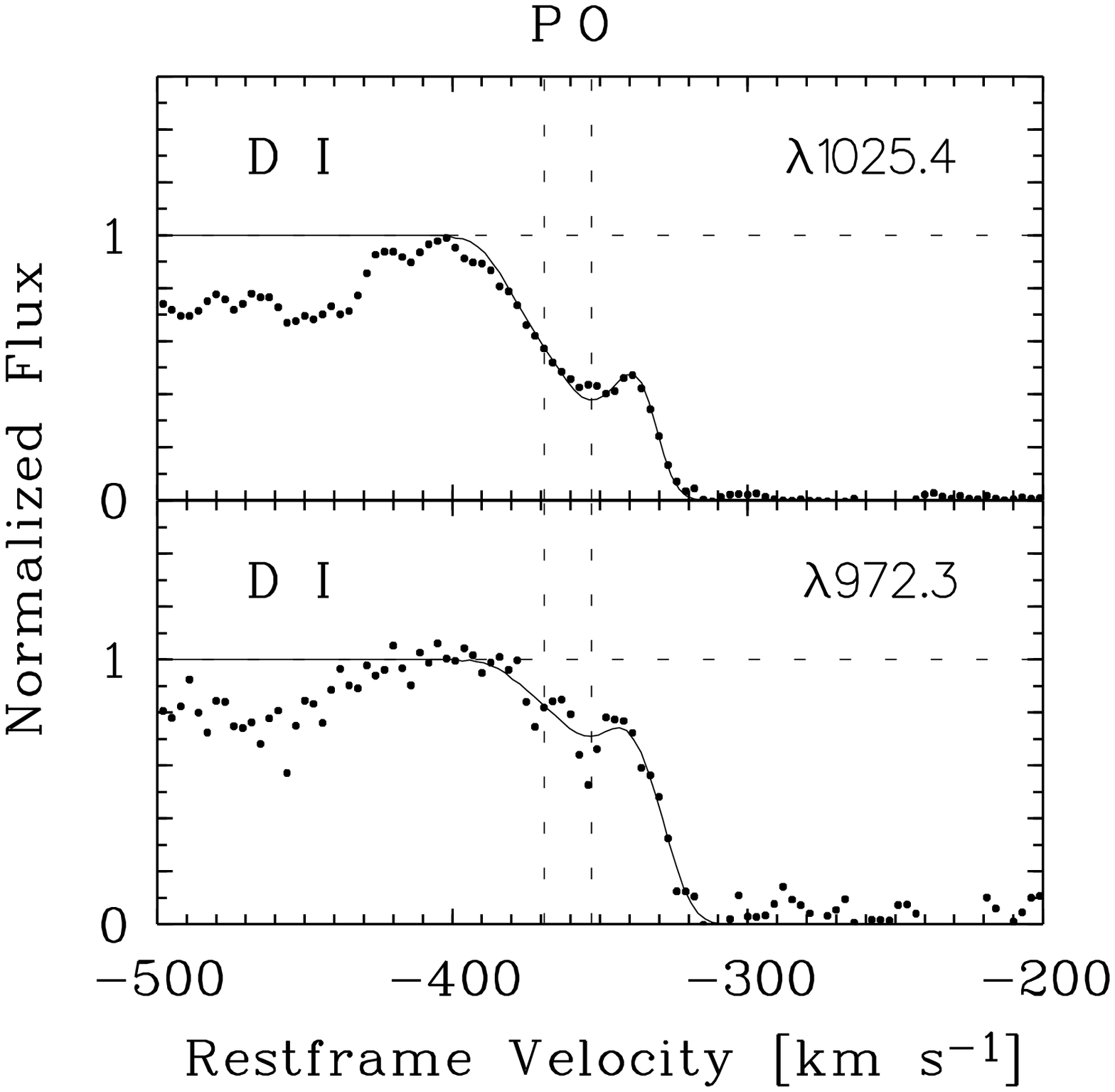}}
\caption[]{
D\,{\sc i} Ly\,$\beta$, and Ly\,$\gamma$
absorption in the $z=2.187$ absorber. The solid line
represents the best fit to the data with log
$N$(D\,{\sc i}$)=13.94\pm0.04$ for component O
and $13.54\pm0.05$ for component P assuming
$b=10.8\pm1.7$ km\,s$^{-1}$ and $11.6\pm1.9$ km\,s$^{-1}$,
respectively.
}
\end{figure}

The H\,{\sc i} profile fit described in
Sect.\,3.3 (Fig.\,5) suggests the
presence of additional absorption in the blue
wing of the Ly\,$\alpha$, Ly\,$\beta$, and Ly\,$\gamma$
profiles - exactly at the position where
D\,{\sc i} absorption from the
two bluer-most components O and P is
expected. A Voigt-profile fit to
the possible D\,{\sc i} Ly\,$\beta$ and Ly\,$\gamma$ absorption
(shown in Fig.\,A.1)
provides log $N$(D\,{\sc i}$)_{\rm O}=13.94\pm0.04$ and
log $N$(D\,{\sc i}$)_{\rm P}=13.54\pm0.05$ together
with $b$(D\,{\sc i}$)_{\rm O}=10.8\pm1.7$ and
$b$(D\,{\sc i}$)_{\rm P}=11.6\pm1.9$. The values for
log $N$(D\,{\sc i}) do not change significantly 
if one assumes lower $b$ values. 
With $N$(H\,{\sc i})$_{\rm OP}\leq 18$ (see Sect.\,3.3),
we derive a limit for the deuterium-to-hydrogen
ratio of (D/H)$_{\rm OP}\geq1.22\times10^{-4}$. This
is $\sim 4$ times higher than the
canonical value for chemically young gas
($3.0\times10^{-5}$; e.g., Burles et al.\,2001).
The simplest and most likely explanation for this
unusually high (D/H) ratio is to assume
that the deuterium lines in components
O and P are blended with an `hydrogen interloper',
a weak H\,{\sc i} absorber that would be located
near $-440$ km\,s$^{-1}$ (in the $z=2.187$
restframe) and
that coincidentally falls together with the
D\,{\sc i} absorption in
components O and P.
However, as recently
reviewed by Jedamzik (2002), there
are a number of possible production
mechanisms at high $z$ that may {\it locally} enhance
the (D/H) ratio in certain environments,
e.g., in regions predominantly
enriched by supermassive stars. It 
is an intriguing coincidence that 
the apparently enhanced (D/H) ratio
in the $z=2.187$ absorber toward
HE\,0001$-$2340 is observed in gas that
exhibits abundance anomalies also in
other elements (P, Si, and Al; see Sect.\,4.2).
We therefore have to acknowledge that the existence of an intrinsically
enhanced (D/H) ratio in components O and P cannot be
entirely excluded, although the presence of an
hydrogen interloper certainly remains the most likely 
explanation for the apparent (D/H) anomaly.

\end{document}